\documentclass[12pt,preprint]{aastex}
\usepackage[yyyymmdd,hhmmss]{datetime}
\usepackage[margin=1.0in]{geometry}
\usepackage{amsmath}
\usepackage{gensymb}
\usepackage{natbib}

\shorttitle{Red Supergiants in M31}

\slugcomment{Accepted by the Astronomical Journal}

\shortauthors{Massey \& Evans}

\begin{document}

\title{The Red Supergiant Content of M31\altaffilmark{*}}

\author{Philip Massey\altaffilmark{1, 2} \and Kate Anne Evans\altaffilmark{1, 3}}

\altaffiltext{*}{Observations reported here were obtained at the MMT Observatory, a joint facility of the University of Arizona and the Smithsonian Institution. This paper uses data products produced by the OIR Telescope Data Center, supported by the Smithsonian Astrophysical Observatory.}
\altaffiltext{1}{Lowell Observatory, 1400 W Mars Hill Road, Flagstaff, AZ 86001; kevans@caltech.edu; phil.massey@lowell.edu.}
\altaffiltext{2}{Department of Physics and Astronomy, Northern Arizona University, Flagstaff, AZ, 86011-6010.}
\altaffiltext{3}{Research Experience for Undergraduate participant during the summer of 2015.  Current address: California Institute of Technology, 1200 East California Blvd, Pasadena, CA 91125; kevans@caltech.edu. }

\begin{abstract}

We investigate the red supergiant (RSG) population of M31, obtaining radial velocities of 255 stars.  These data substantiate membership of our photometrically-selected sample, demonstrating that Galactic foreground stars and extragalactic RSGs can be distinguished on the basis of {\it B-V, V-R} two-color diagrams.  In addition, we use these spectra to measure effective temperatures and assign spectral types, deriving physical properties for 192 RSGs.  Comparison with the solar-metallicity Geneva evolutionary tracks indicates astonishingly good agreement.  The most luminous RSGs in M31 are likely evolved from 25-30$M_\odot$ stars, while the vast majority evolved from stars with initial masses of 20$M_\odot$ or less.
There is an interesting bifurcation in the distribution of RSGs with effective temperatures that increases with higher luminosities, with one sequence consisting of early K-type supergiants, and with the other consisting of M-type supergiants that become later (cooler) with increasing luminosities. This separation is only partially reflected in the evolutionary tracks, although that might be due to the mis-match in metallicities between the solar Geneva models and the higher-than-solar metallicity of M31.  As the luminosities increase the median spectral type also increases; i.e., the higher
mass RSGs spend more time at cooler temperatures than do those of lower luminosities, a result which is new to this study. Finally we discuss what would be needed observationally to successfully build a luminosity function that could be used to constrain the mass-loss rates of RSGs as our Geneva colleagues have suggested. 
\end{abstract}


\vskip 30pt

\section{Introduction}
\label{Sec-intro}

Massive stars spend most of their lives as OB stars.  The most massive of these ($>\sim$40$M_\odot$) then
evolve into Wolf-Rayet (WR) stars, possibly after passing through a luminous blue variable (LBV) phase. By contrast, massive stars with masses below $\sim 30M_\odot$ spend their He-burning lives as red supergiants (RSGs) after passing through a yellow supergiant (YSG) stage during their quick journey across the H-R diagram.  At some  intermediate masses ($\sim 30 M_\odot$, say) stars may pass through both a RSG and WR phase, passing through the YSG phase twice.  
The exact mass ranges corresponding to these various stages, along with the relative lifetimes spent in these regions of the HRD, depend heavily on 
the initial metallicity of the gas out of which these stars form, as radiation pressure acting on highly ionized
metals results in mass loss that strongly influences this evolution.  Thus, characterizing the luminous populations of nearby galaxies allows us to perform exacting tests of massive star evolution as a function of metallicity. (For a recent review, see \citealt{MasseyRev13}.)

The Andromeda Galaxy (M31) plays a unique role in such studies, as it provides the only nearby extragalactic example where the metallicity is solar and above \citep{Zaritsky,Sanders}.  Previous work on the massive star population of M31 has established the WR content \citep{NeugentM31}, identified YSGs \citep{DroutM31}, discovered LBV candidates \citep{M31PCyg,LGGSIII,BigTable},  and laid the  groundwork for the current study by performing a preliminary reconnaissance of its RSG population \citep{MasseyRSGs,MasseySilva}.  Work on the unevolved massive star population suggests that only a few percent
of the total number O stars and B supergiants have been identified \citep{BigTable}.  

\begin{figure}
\epsscale{0.8}
\plotone{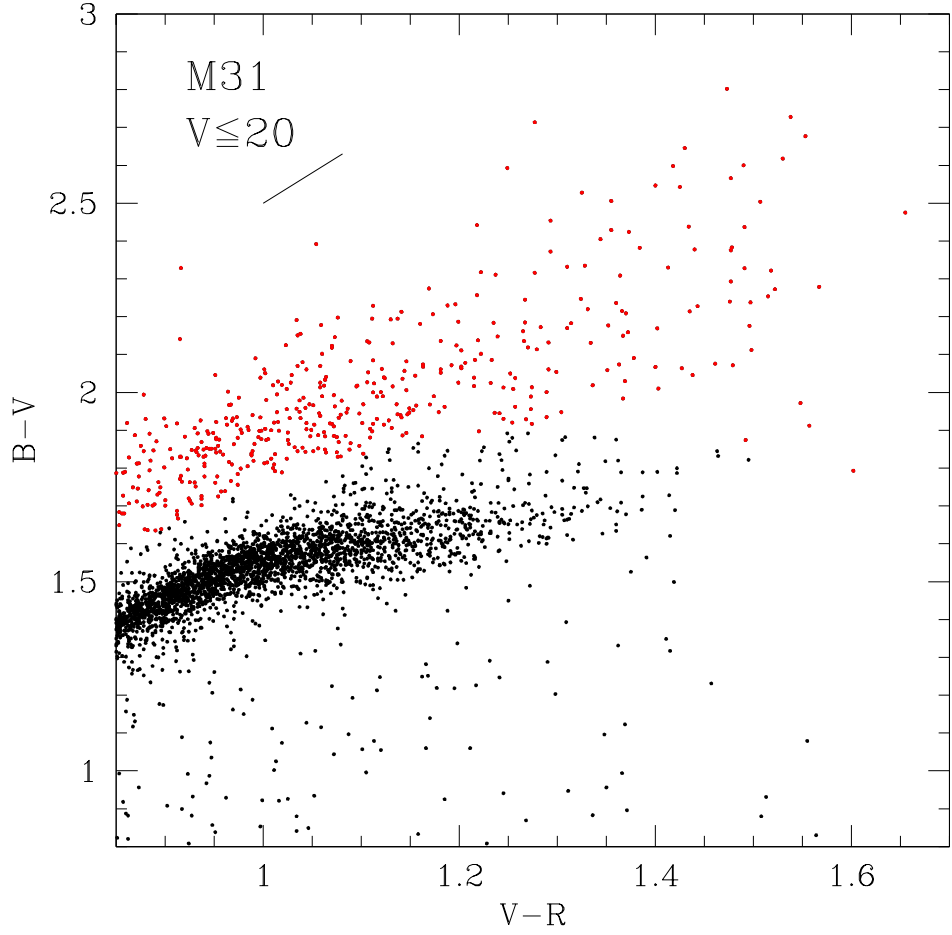}
\epsscale{1.0}
\caption{\label{fig:m312col} Two-color diagram of red stars in M31. The points in red are suspected red supergiants from Table~1 of \citet{MasseySilva}, while the points in black represent suspected foreground stars.  The photometry is taken from the LGGS \citep{LGGSI}, with $V\le20.0$.  The typical reddening vector for M31 stars is shown by the diagonal line. }

\end{figure}

In this paper we set out to complete our identification of the RSG population of M31 down to $\sim$15$M_\odot$,
 using a combination of spectroscopy and photometry.   \citet{LGGSII} showed that samples of red stars in the right color and magnitude range to be RSGs were badly contaminated ($\sim$80\%) by foreground red dwarfs.  However, for
cool stars ($T_{\rm eff}\le4300$ K), {\it B-V} becomes primarily an indicator of surface gravity due to line-blanketing in the {\it B} band, while {\it V-R} remains primarily a temperature indicator \citep{MasseyRSGs}.
Thus there is a relatively clean separation in a {\it B-V} vs {\it V-R} two-color plot between foreground dwarfs and RSGs.
We show an example of such an effort in  Figure~\ref{fig:m312col} based upon \citet{MasseySilva}.

Prior efforts at spectroscopic confirmation have been relatively modest.  \citet{MasseyRSGs} ``confirmed" about 20 RSGs
in M31.  In some cases the spectroscopic confirmation is based on clear radial velocity information, but in other cases it is based on softer criteria, such as the strengths of the Ca II triplet lines; these criteria were not always in good agreement. \citet{MasseySilva} identified a sample of 437 RSG candidates in M31 from the Local Group Galaxy Survey (LGGS) {\it BVR} photometry of \citet{LGGSI}, and obtained radial velocities of 124 of these stars, but derived physical properties (effective temperatures and bolometric luminosities) for only 16 stars in their sample.  Furthermore, the photometrically selected RSG and foreground candidates did not always prove to be what was expected when observed spectroscopically.  We decided it was time for a more complete study. 

This investigation is timely.  As the evolutionary models improve, knowledge of the RSG content of M31 will provide us with a powerful magnifying glass\footnote{\citet{Kipp} liken  these evolved stages to ``a sort of magnifying glass, [which reveals] relentlessly the faults of calculations of earlier phases."  We are indebted to our colleague Andre Maeder for calling this quote to our attention.} for examining the predictions of these models, and for better constraining the input.  For instance,
\citet{Maeder80} argued that the relative number of RSGs and WRs should be a strong function of the metallicity,
with proportionately fewer RSGs found at higher metallicities.  Further, the number of RSGs gives another way of estimating the expected number of unevolved O stars, a value that is poorly constrained observationally \citep{BigTable}.

An even more exciting possibility is for us to use our knowledge of the RSG content of M31 to determine the mass-loss rates for RSGs.   As recently shown by \citet{GeorgesRSGs} the time-averaged mass-loss rates of RSGs are very poorly constrained by observations, as much of the action occurs episodically.  
The consequences of this significantly affect our interpretation of the populations of the upper part of the HRD.  For instance, according to the models, enhanced mass-loss during the RSG phase leads to the prediction
that the {\it majority} of blue and yellow supergiants are post-RSG objects \citep{GeorgesRSGs}.  In addition, enhanced mass-loss rates during the RSG phase may solve the so-called ``red supergiant problem."  RSGs have long been thought to be the progenitors of type II-P supernovae; \citet{SmarttIIP}, however, argued that 
the observed upper mass limit of the progenitors of type II-P SNe is 17-18$M_\odot$ rather than the 30$M_\odot$ or more one would expect from what the models.  \citet{Sylvia} argues that an improved mass-loss prescription during the RSG phase leads naturally to this result, as the mass-loss rates are then significantly enhanced for the highest mass RSGs, leading to a likely scenario where these stars lose most of their hydrogen-rich envelopes and exploded instead as type II-L or even type Ib supernovae. 

The validity of this resolution rests on the assumption that the RSG mass-loss rates have been seriously underestimated.  \citet{GeorgesRSGs} and \citet{CyrilIAU} proposed a novel way of determining if this is correct.   The evolution models predict that the luminosity functions of RSGs will have a strong dependence on the average mass-loss rates for RSGs,
becoming steeper (fewer high luminosity RSGs) with higher mass-loss rates.  However, this requires an unbiasd knowledge of the RSG content of a mixed-age population in which the (global) star-formation rate has remained essentially constant over the relevant evolution time period (20-30~Myr).    Model calculations are currently underway for this undertaking.  Our study here does not provide an answer, but it provides an vital toe-in-the-door that delineates the observational hurdles associated with constructing the required bias-free luminosity function.

In Section~\ref{Sec-obs} we will explain our sample selection and the spectroscopy we obtained for our study. In Section~\ref{Sec-Analysis} we will use these data for our analysis, first distinguishing RSGs from foreground stars using the radial velocities (Section~\ref{Sec-membership}), and comparing these results to our expectations based on our two-color diagrams.  We will also measure effective temperatures and spectral types (Section~\ref{Sec-temps}), and compute other physical properties, such as the bolometric luminosities.  We will then compare these to what the Geneva evolutionary tracks predict (Section~\ref{Sec-HRDs}).  Finally, in Section~\ref{Sec-discussion} we will summarize our results and discuss what future work is needed in order to construct a sufficiently deep luminosity function for constraining the RSG mass-loss rates.

\section{Observations}
\label{Sec-obs}
\subsection{Sample Selection} 

Our primary goal was to see how accurately we could  use the ({\it V-R, B-V}) two-color selection method of \citet{MasseyRSGs} to distinguish actual RSGs from Galactic foreground stars.  In this way, M31 serves as an excellent (high-metallicity) test bench.
For some galaxies, such as the Magellanic Clouds, most of the systemic radial velocity is actually due to the reflex
motion of the sun.  Thus, although foreground disk dwarfs are easily distinguished from Magellanic Cloud members
\citep{NeugentSMC,NeugentLMC}, halo red giants would be hard to distinguish from Magellanic Cloud RSGs on the basis of radial velocities alone.  (Statistically, however, we can argue that few halo red giants are expected
in the appropriate magnitude range; see Section 4 of \citealt{NeugentLMC}.)  M31 has a systemic velocity of $\sim$$-300$~km~s$^{-1}$ and a rotational velocity of $\sim$250~km~s$^{-1}$ (\citealt {MasseySilva} and references therein.)
Thus there is an ``alligator's jaws" shaped area in the  NE section of M31 where it is difficult to separate foreground dwarfs from M31 members, but over most of the galaxy the separation is very clean. (See Figure~5 of \citealt{DroutM31}.) 

In selecting targets to observe, we used the 437 photometrically selected RSG candidates
given in Table~1 of \citet{MasseySilva}.  These are the stars marked as red in Figure~\ref{fig:m312col}, and met the
following criteria: (1) $V<20$, corresponding roughly to $\log L/L_\odot \sim 4.4$ for early K-type stars, and $\log L_/L_\odot \sim 4.8$ for
mid M-type stars.  (2) $V-R\ge0.85$ (and correspondingly $B-V\ge 1.5$) to restrict the sample to K-type stars and later, roughly corresponding to $T_{\rm eff}<4300$ K given typical reddening for M31 RSGs\footnote{Our interpretation of the $V-R$ cutoff requirements differs slightly from that stated by \citet{MasseySilva}, who state that $V-R\ge0.85$ corresponds to an intrinsic $(V-R)_0$ of 0.81 ($T_{\rm eff} \leq$4000~K).  Yet, we expect $E(V-R)\sim0.6E(B-V)$ \citep{Sch}.  The M31 OB stars in the LGGS have a median reddening of $E(B-V)=0.13$, so this would corresponds to an $E(V-R)=0.77$, not 0.81.  But, the actual M31 RSG sample has higher reddening, with $E(B-V)\sim0.3$ for reasons probably associated with circumstellar dust \citep{Smoke}.  Thus, $E(V-R)\sim0.18$, and $(V-R)_0 \sim 0.67$, corresponding more to an $T_{\rm eff}\leq$4300~K.  This allows early-type K stars to be included in the sample.}.   (3) $B-V>-1.599(V-R)^2+4.18(V-R)-0.83$ to separate the low surface gravity RSGs from the high surface gravity foreground Galactic stars.  Some of these stars had previously measured radial velocities \citep[Table 2 of][]{MasseySilva},  which could then serve as radial velocity templates.    Of the 437 stars, 36 (8.2\%) of them were somewhat crowded, and given lower priorities in the Hectospec assignments.  

This project was ``piggy-backed" on our primary program to continue to monitor a sample of WR binaries in M31 \citep{NeugentBinaries}.  We observed six of these WR configurations with slightly different supergiant and foreground candidate red stars included on the otherwise unused fibers.  In addition, we observed one configuration that consisted only of red stars.  In all we successfully observed 255 RSG candidates and 98 foreground candidates, many of them multiple times.

\subsection{Spectroscopy}
The observations were all made with Hectospec, a 300-fiber spectrometer on the 6.5-m MMT \citep{Hecto}.   The 270~lines~mm$^{-1}$ grating was used, providing a reciprocal dispersion of 1.2~\AA\ pixel$^{-1}$, and a spectral resolution of
4.5-5.2~\AA\ from 3650-9200~\AA\ in first order.  Since no
blocking filter is used, there is some contamination from second-order blue in the far red, but given the colors
of these stars, this is pretty minimal; the grating is blazed at 5000~\AA.    The six ``WR configurations" were observed
for 90 minutes each (UT 2014 Sep 24, Sep 26, Nov 21, Nov 22, Nov 26, and Nov 28); the one ``red star" configuration was observed for 60 minutes (UT 2014 Oct 1).  The observations were made in a self-staffed queue; P. M. and collaborator Kathryn Neugent were present for the November observing.    

The reductions were previously described by \citet{KateRunaway}. 
For each of the seven configurations, some fibers were assigned to blank sky in order to facilitate sky subtraction.
Calibration exposures included HeNeAr and quartz lamp exposures.  The data were all run through the SAO pipeline, with the night-sky lines used to adjust the wavelength zero-points relative to the HeNeAr exposures
that were made in the afternoon.   As part of the reduction procedure, the wavelengths were corrected to the
heliocentric rest frame.  Observations of the spectrophotometric standard Feige 34 were made midway through the semester, and were used to produce sensitivity curves by Nelson Caldwell, who kindly made these available to us.  

\section{Analysis}
\label{Sec-Analysis}

\subsection{Membership: Red Supergiants vs Foreground Stars}
\label{Sec-membership}

We measured radial velocities from each of the spectra using {\sc xcsao}, an IRAF\footnote{IRAF is distributed by the National Optical Astronomy Observatory, which is operated by the Association of Universities for Research in Astronomy (AURA) under a cooperative agreement with the National Science Foundation.} cross-correlation tool.   As described in
\citet{KateRunaway}, we used 21 Hectospec spectra of six M31 RSGs for which \citet{MasseySilva} had already determined radial velocities. (Those velocities, in turn, were tied to three legitimate late-type radial velocity standards.) 
The cross-correlation was restricted to the 8350-8750~\AA\ region around the Ca\, {\sc ii} $\lambda \lambda 8498, 8542, 8662$ triplet, lines that are very strong in RSGs.  This process also allows us to avoid the wide molecular features, atmospheric bands, and so on.  For each individual program spectrum, a single velocity was produced by averaging the results from the 21 velocity templates, weighting each appropriately by the internal  error from 
the cross-correlation (i.e., weights were the squares of the reciprocal errors).  For the stars with multiple observations,
a weighted average was then determined\footnote{Prior to 2014, the SAO  pipeline reductions did {\it not} include correcting the
wavelength scale to heliocentric; this change caused a certain amount of confusion and consternation during our
initial attempts to understand our results.  We are indebted to Nelson Caldwell for help in tracking down
this issue.}.  We list the radial velocities in Table~\ref{tab:RVs}.

\citet{MasseySilva} demonstrated that the radial velocities of M31's HII regions and RSGs agreed 
very well, and that the expected radial velocities of these Population I objects
could be approximated by a linear relationship with $(X/R)$,  
$$V_r = -295 + 241.5 (X/R),$$ where $X$ is the distance along the semi-major axis, and $R$ is the galactocentric distance within the plane of M31.  Such a linear approximation agrees well with the more complex two-dimension velocity field determination of \citet{M312D} and other studies \citep[e.g.,][]{Hurley}, and is equivalent to saying that the rotation curve is flat.  By contrast, we expect the foreground red stars to cluster around a velocity
of 0 km s$^{-1}$, with some scatter indicative of the radial velocity dispersion of nearby red dwarfs in this particular
direction.  

\begin{figure}
\epsscale{0.8}
\plotone{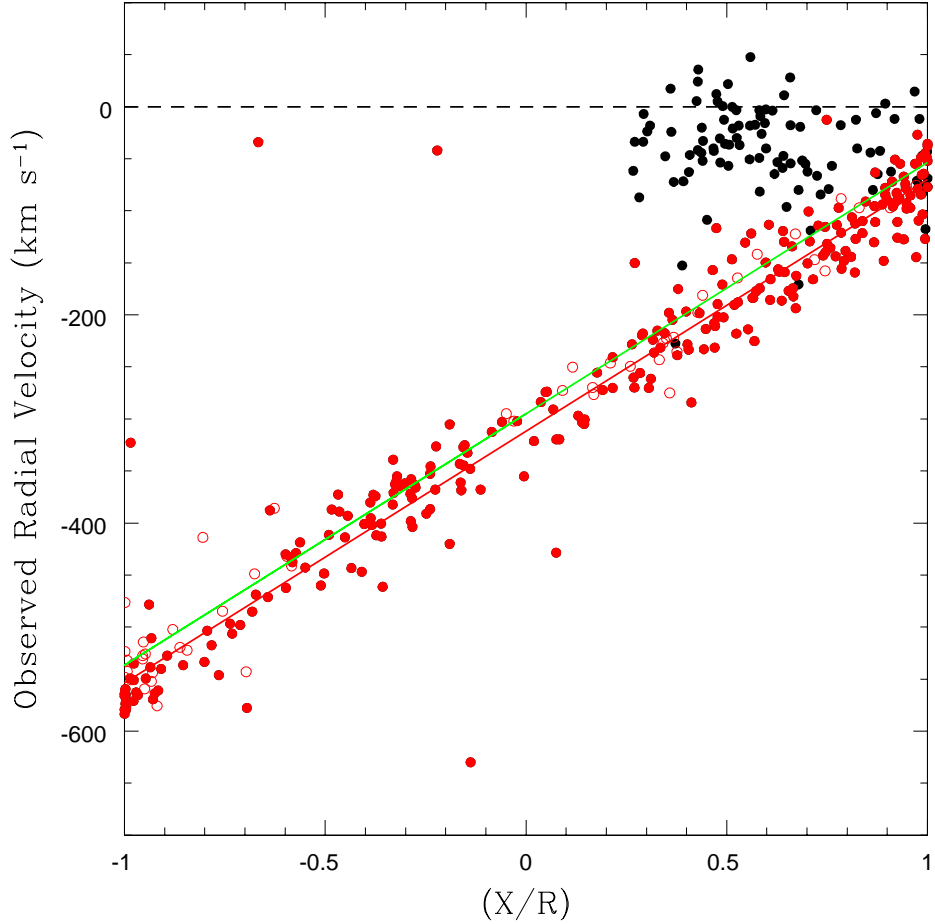}
\epsscale{1.0}
\caption{\label{fig:rvplot} Radial velocities of foreground and red supergiant candidates. The radial velocities of our foreground (black) and RSG (red) candidates are plotted against $(X/R)$, where $X$ is the distance along the semi-major axis, and $R$ is the galactocentric separation within the plane of M31.   Filled red circles are from the present paper; open red circles are from \citet{MasseySilva}.  The black dashed line indicates 0 km s$^{-1}$, and we expect that foreground galactic stars will cluster near this value.  The green line denotes the radial velocity vs $(X/R)$ relation
for M31 from \citet{MasseySilva}, while the red line denotes the revised version given here.}

\end{figure}

In Figure~\ref{fig:rvplot} we show a plot of the radial velocity with $(X/R)$ for all of our data.  We see that the vast majority of the candidate RSGs (red points) follow the expected relationship between $V_r$ and $(X/R),$ while the vast majority
of the candidate foreground stars (black points) cluster around 0 km s$^{-1}$ as expected.  The foreground candidates were observed only as part of the  single ``red star" configuration, and hence are distributed over the limited range
of $(X/R) \sim 0.25$ to 1.0.  We have used the larger
sample size here to improve on the radial velocity relation, deriving $$V_r = -311.8 + 242.0 (X/R).$$  Over the range $(X/R)=-1$ to +1, the expected $V_r$ from this relationship differs from that of the \citet{MasseySilva} determination by $-16.3$ to $-17.3$ ~km~s$^{-1}$.  This may be compared to the overall scatter around the relationship of 25 km$^{-1}$. Recall that the radial velocity ``standards" for the new measurements came from adopting the radial velocities determined by \citet{MasseySilva}.   We checked our new measurements against the old to make sure this $\sim$$-17$ km s$^{-1}$ was due to better coverage in $(X/R)$ and not some reduction issue.  Indeed, there is only a small systematic offset between the
new and old measurements; the mean difference (in the sense of new {\it minus} old)   is $-7.4$ km s$^{-1}$ with a scatter of 12.3 km s$^{-1}$, and median difference $-5.9$ km s$^{-1}$ using 73 stars in common\footnote{This excludes one outlier with a huge difference:   J004217.99+410912.7 whose velocity measurement here is 118 km s$^{-1}$ more positive than that in \citet{MasseySilva}.  Neither measurement agreed well with expected radial velocity; possibly the star is a radial velocity variable (binary).}.   If we include the 50 additional RSGs with radial velocities from \citet{MasseySilva} for which we do not have new radial velocities, we obtain a very similar relation, $$V_r =-310.0+239.7(X/R).$$

Although our photometric classification was obviously extremely successful, we can see that there were a few
RSGs misclassified photometrically as foreground stars, a few  foreground stars misclassified as RSGs, and a few
RSGs with interestingly discrepant radial velocities.  Let us consider each of these cases in turn.

\begin{figure}
\epsscale{0.8}
\plotone{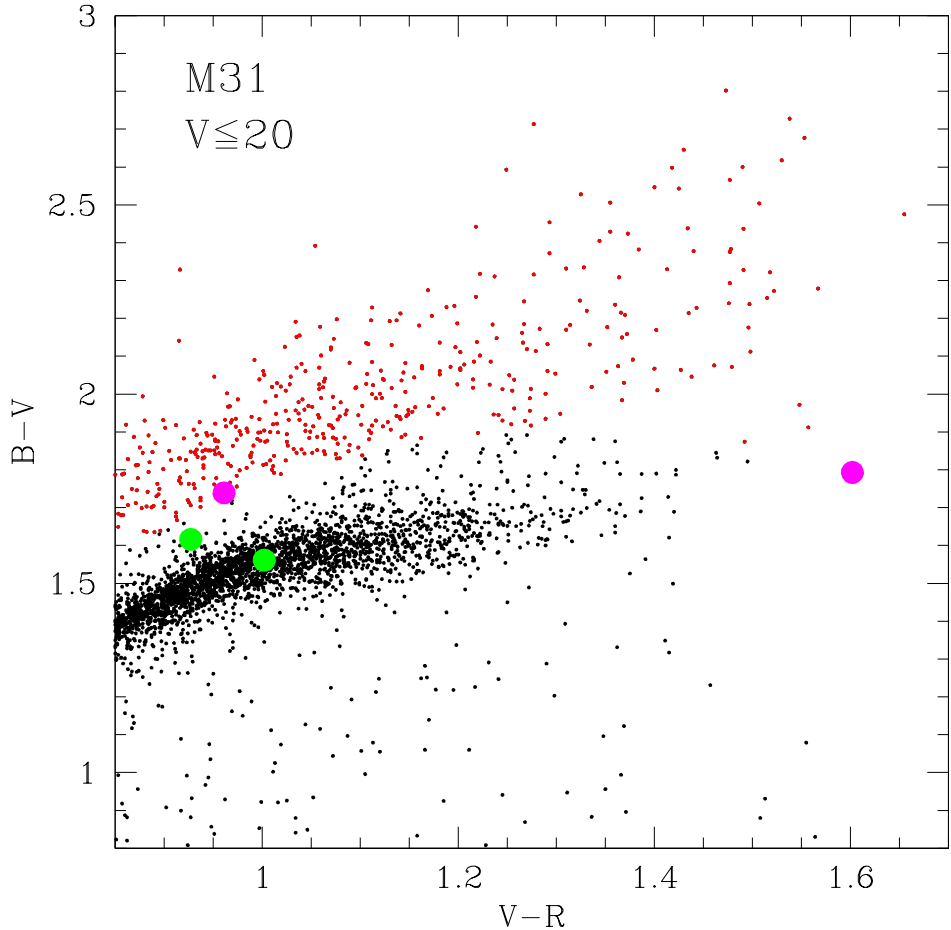}
\epsscale{1.0}
\caption{\label{fig:weirdos} Location of discrepantly classified stars (from Figure~\ref{fig:rvplot}) in a two-color diagram.  Same as Figure~\ref{fig:m312col} but with large symbols denoting the location of the four stars whose radial velocities disagree
with the photometric classification.  Green symbols are stars photometrically classified as foreground objects but
whose radial velocities are consistent with M31 membership, while magenta symbols are stars photometrically classified as RSGs, but whose radial velocities are consistent with membership in the Milky Way.  A fifth star, whose photometry is confused by crowding, is located off the right side of this plot.}
\end{figure}

\citet{DroutM31} argued that confusion with foreground stars sets in at radial velocities $\geq$-150~km~s$^{-1}$.  
There are two foreground candidates (black points)  more negative than that in Figure~\ref{fig:rvplot}, each near the RSG velocity
relation: in other words, two stars whose velocities indicate they are indeed RSGs despite having been classified
photometrically as foreground stars.  These are J004458.49+421219.2  and J004406.92+412307.2.   We denote their position in the two color diagram in Figure~\ref{fig:weirdos} with large green points. We see that one of these stars
(J004458.49+421219.2) has  photometry right on the border between what we defined as a RSG candidate and what we called a foreground candidate; the other,  J004406.92+412307.2, really ``should" be a foreground star according to the photometry. 

There are three stars photometrically classified as RSGs that have velocities consistent with their actually being foreground stars: J004105.97+403407.9, J004303.26+404710.9, and J004431.71+415629.1.  We mark two of these stars
with large magenta symbols in Figure~\ref{fig:weirdos}; the third (J004105.97+403407.9) has such unrealistic photometry ($V-R\sim 4.7$) that it would fall well to the right on the plot.  That star is extremely crowded according to the LGGS.  The other two stars have photometry that is right on the borderline between the separation of RSGs and foreground stars.  Of these, the one with the redder {\it V-R}, has very poor spectra, as well as a surprisingly large {\it V-R} color.   

Thus, of all 354 stars with new data, only one star is surprisingly inconsistent with its photometric classification vs.\ its radial velocity.   We conclude that the photometric classification using the {\it V-R, B-V} two color diagram appears to be a very robust way of separating foreground stars from RSGs without needing radial velocities, at least at high metallicities.

The final type of discrepancy is when a RSG candidate has a surprising velocity with respect to its position in M31, but its velocity is not consistent with it being a foreground star either. 
\citet{KateRunaway} discuss {\bf J004330.06+405258.4,} the star with a radial velocity of $-630$~km~s$^{-1}$ at $(X/Y)=-0.138$.   They conclude that this is an bona fide runaway RSG, the first discovered on the basis
of its radial velocity, and the first known extragalactic massive star runaway.  A significant percentage (10-50\%) of unevolved massive stars are OB runaways \citep[][and references therein]{Gies}, with discrepant radial velocities ($>$30~km~s$^{-1}$) compared to other OB stars in their neighborhood.  Runaways probably occur via dynamical
evolution \citep{Gies,fujii}.  Few evolved massive runaways are known, doubtless because of the difficulties in recognizing such objects once they are older and have moved further from their birthplace.  J004330.06+405258.4 has a radial velocity that is 300~km~s$^{-1}$ more negative than that expected from M31's rotation curve, and \citet{KateRunaway} note that the is located some $\sim$20\arcmin\ (4.6~kpc) from the plane of the disk, consistent with it having a transverse velocity similar to its radial velocity motion.  

We find three other potential runaways in Figure~\ref{fig:rvplot}: {\bf J004217.99+410912.7}  has a radial velocity
of $-323$~km~s$^{-1}$ rather than the $-533$~km~s$^{-1}$ expected for $(X/R)=-0.985$.  We have five observations of this star, all with similar
velocities.  However, an older measurement by \citet{MasseySilva} is disagrees with these, with a value of $-440$
~km~s$^{-1}$.   The difference in radial velocity between the Fall 2014 observations reported here and the Fall 2005
observations of \citet{MasseySilva} might suggest the star is a long-period binary, but the large velocity difference would be more indicative of a short-period binary, which we can essentially rule out from our five new measurements.  These results are puzzling.  

{\bf J004112.27+403835.8} and {\bf J004228.99+412029.7} can be seen in Figure~\ref{fig:rvplot} at ($-0.695$, $-577$~km~s$^{-1}$) and ($0.075$, $-428$~km~s$^{-1}$).  These stars also might be runaways as their radial velocities are expected to be $-462$~km~s$^{-1}$ and $-277$~km~s$^{-1}$, respectively.  Both stars have multiple observations, with good agreement in their measurements, as evidenced by the low standard errors of the means given in Table~\ref{tab:RVs}, i.e., 1.0 ~km~s$^{-1}$and 1.3~km~s$^{-1}$. However, the spectrum of J004228.99+412029.7 is that  of a very early K, or even late G-type; this is consistent with the low value for the 
\citet{TonryDavis} $r$ parameter we obtained, indicating a relatively poor match with the M-type radial velocity templates we used in measuring the velocities.

\subsection{Effective Temperatures, Spectral Types and Physical Parameters}
\label{Sec-temps}

Spectral types for the M31 RSGs were determined by comparing their spectra to those of Galactic RSGs observed by \citet{EmilyMW}.   Those Galactic RSGs had spectral types that were
generally well established in the literature, although in a few cases they were reclassified by \citet{EmilyMW}.   The agreement between these were usually exact, although occasionally there would be a difference of one one spectral subtype,
i.e.,  M1~I vs.\ M2~I.   The primary criteria were those described by \citet{EmilyMW}, i.e., the strengths of the TiO bands for M- and late K-types, and the strengths of the G band and 
Ca\, {\sc i} $\lambda 4226$.  We list these spectral types in Table~\ref{tab:physical}.  In the few cases where no spectral type could be assigned, the stars are likely not RSGs at all, but are somewhat earlier members of M31. 

Effective temperatures were determined following the procedures described by \citet{EmilyMW} using the $2\times$ solar metallicity MARCS atmosphere models used by
\citet{MasseySilva}.   For the vast majority of our M31 RSGs we had multiple observations; each
of these spectra were fit independently; what are shown are the median values.  The uncertainty in our fitting is approximate 50~K. These temperatures are given in Table~\ref{tab:physical}.  Note that in a few cases effective temperatures could not be measured reliability.

There are eight stars in common between our study here and that of \citet{MasseySilva}; we list a comparison of the spectral types and effective temperatures in Table~\ref{tab:comparison}.  In preparing this we were chagrined to discover a significant error in \citet{MasseySilva}: the effective temperatures they listed were based upon the solar-metallicity MARCS models and not the 2$\times$ solar metallicity MARCS model as stated in the text.  Experimentation showed that this introduced a systematic difference of 75 K; i.e., that fitting the spectra used by \citet{MasseySilva} with the $2\times$ solar model rather than the $1\times$ solar model require temperatures that are roughly 75 K higher.
We have therefore adjusted the \citet{MasseySilva} effective temperatures by that amount in making this comparison.
We find that the agreement in the effective temperatures determined from our new spectra to be quite consistent.
The only significant discrepancy in spectral type is for J003957.00+410114.6.  We have six observations of that star; all yield a spectral type in the range of M3-M4~I, consistent with our 
effective temperature.  We are forced to conclude that the \citet{MasseySilva} spectral type is a mistake rather than
a sign of variability, given that the effective temperature given for the star by \citet{MasseySilva} is much cooler
than what they found for other M0~I stars, and is more like that of a M3~I type.   

In the past, our standard procedure \citep[e.g.,][]{EmilyMW,EmilyMC,EmilyWLM,MasseySilva} has been to use the model fitting to determine $E(B-V)$.  However, our recent experiences with fiber spectrophotometry has been disappointing.   Hectospec uses an atmospheric dispersion compensator, and a careful study by \citet{FluxHecto} demonstrates that the flux calibration of Hectospec data is good to a few percent on extended sources.  However, we have found that flux calibrating point sources in crowded fields is considerably less successful.  In many cases the reddenings required to fit the fluxed spectrophotometry are clearly non-physical, i.e., requiring zero or negative $E(B-V)$s.  We have encountered similar problems in modeling Magellanic Clouds RSGs using 2 Degree Field fiber data from the Australian Astronomical Observatory in a separate project.   We have decided then to adopt the median $E(B-V)=0.3$ ($A_V=1.0$) found by \citet{MasseySilva} for their sample of M31 RSGs.  This is somewhat larger than the average reddenings of OB stars \citep{LGGSII}, which is consistent with what we find from a comparison the reddenings of OBs and RSGs in the Milky Way \citep{EmilyMW} and Magellanic Clouds \citep{EmilyMC}, and which is attributable to circumstellar reddening by dust \citep{Smoke}. 

To minimize the effect that $A_V$ is poorly determined individually for our stars, we have chosen to rely on NIR $K$-band photometry, as $A_K$ is only about 10\% of $A_V$, and the bolometric correction is small, and fairly insensitive to the temperature. 
  This also has the advantage of avoiding the typical $\sim 0.9$~mag variability seen in $V$ \citep{EmilyVariables}; no such variability is seen at $K$, a point also emphasized by \citet{MasseySilva}.  Of the 255 stars in our sample, 37 have 	targeted K-band photometry using the NIR photometer FLAMINGOS \citep{MasseySilva}; for the rest, we adopt 2MASS \citep{2MASS} values where available.  We perform the same conversion to bolometric luminosity in the same manner as \citet{MasseySilva}: 
  $$K=K_s+0.04$$
  $$K_0=K-0.12A_V=K-0.12$$
  $$M_K=K_0-24.40$$
  $$M_{\rm bol}=M_K+{\rm BC}_K,$$
  where K is the ``standard" (such as it is) photometric band \citep[see, e.g.,][]{Carpenter}, $K_s$ is the observed 2MASS or FLAMINGOS photometry, $K_0$ is the de-reddened $K$-band magnitude, $M_K$ is the absolute $K$-band magnitude where we have assumed a true distance modulus of 24.40 \citep[0.76 Mpc,][]{vandenbergh2000}, $M_{\rm bol}$ is the bolometric magnitude, and ${\rm BC}_K$ is the bolometric correction in the K-band, a small but positive number \citep[e.g.,][]{Bessell,EmilyMW}.  We determine ${\rm BC}_K$ from the effective temperature $T_{\rm eff}$ following \citet{MasseySilva} as follows: $${\rm BC}_K=7.149 - 1.5924(\frac{T_{\rm eff}}{1000~{\rm K}}) + 0.10956(\frac{T_{\rm eff}}{1000~{\rm K}})^2,$$ where the relationship is derived from the MARCS models.  Finally, of course, $$\log L/L_\odot =(M_{\rm bol}-4.75)/ -2.5$$.

\subsection{Comparison with Evolutionary Tracks}
\label{Sec-HRDs}

Having derived physical properties for our sample of RSGs, we compare their location in the H-R diagram (HRD) to that predicted by the evolutionary tracks in Figure~\ref{fig:hrd}.  In making this comparison, we encounter a difficulty: so far, only solar \citep{Sylvia} and sub-solar 
\citep{Cyril002} versions of the new Geneva tracks are available.   \citet{Sanders}  has recently conducted a comprehensive study of the chemical abundances of M31's HII regions, finding that log (O/H)+12 = 9.0 to 9.1 ($2\times$ solar) in the center of M31 with a shallow galactocentric gradient ($-0.01$ to $-0.02$ dex kpc$^{-1}$), in accord with the older study of \citet{Zaritsky}.  (A value of log (O/H) + 12 =  8.7 is considered typical of the solar neighborhood, and hence our description of $2\times$ solar.)  However, one of the most intriguing results to come out of the \citet{Sanders} study
is the large scatter in the oxygen abundance measures, which these authors argue is {\it intrinsic,} and not observational.  As their Figure 8 shows, at a galactocentric distance of 10-12~kpc, the values for log (O/H)+12 vary from 8.5 to 9.3~dex.  Thus, the solar metallicity evolutionary tracks are not inappropriate either.   However, the changes in the physical parameters we would have derived using solar metallicity models are extremely modest.  We would have to decrease our effective temperatures by 75~K, the approximate systematic difference between the solar and $2\times$solar MARCS atmospheric models as discussed earlier.  This translates to a very modest $\sim$0.01~dex decrease in the effective temperature, and 0.02~dex decrease in the bolometric luminosity, i.e., the change is comparable to the point sizes in the figure. 

The agreement between the evolutionary tracks and the locations of stars in the HRD is nothing short of remarkable.  There are only four stars whose luminosities appear to be higher than what the tracks allow. The two highest luminosity stars in the figure are J004520.67+414717.3 (M1~I) and J004539.99+415404.1 (M3~I) are both found at $X/R > 0.9$ where there {\it is} considerable overlap in the radial velocities of RSGs and foreground stars; we thus are relying primarily on their photometry to classify these as supergiant members. The photometry of the first of these clearly places it in the RSG category (J004520.67+414717.3 has $B-V=2.68$, $V-R=1.55$), while the photometry of the second is more marginal (J004539.99+415404.1, $B-V=1.70$, $V-R=0.94$).  The next two most luminous stars are not necessariy discrepant with the evolutionary tracks, as somewhere between 25-32$M_\odot$ may extend over to the RSG phase. These stars have $\log L/L_\odot\sim 5.6$ and are  J004428.12+415502.9 (K2~I, $X/R=0.75$) and J004428.48+415130.9 (M1~I, $X/R=0.82$), located where where there is more support from the radial velocities for membership. However, the photometry of J004428.12+415502.9 has an unusually large $B-V$ value (2.05) for its $V-R$ value (0.95).  Finally, there are three stars all with $\log L/L_\odot$ of 5.45-5.47: J004125.23+411208.9 (M0~I, $X/R=-0.29$),
J004312.43+413747.1 (M2~I, $X/R=0.45$), and J004514.91+413735.0 (M1~I, $X/R=0.64$) where the radial velocities leave no question as to membership; these stars, however, are consistent with the 25$M_\odot$ track.  We conclude that the only J004520.67+414717.3 seems to be discrepant with what the evolutionary tracks predict; this high luminosity star may warrant further investigation.

There is an interesting bifurcation in the location of the RSGs in the HRD, with two branches separated by increasingly large differences in effective temperatures at higher luminosities.  Is this real, or an artifact of our analysis? The warmer stars are those classified as early K's, while the cooler stars are classified as M-type; the larger gap at higher luminosities is a reflection that the median spectral type for the M's increases with luminosity. For instance, for
the stars with $\log L/L_\odot$ less than 5.0, the median spectral type is M0~I; for stars with $\log L/L_\odot\ge5.0$,
the median spectral type is M2~I.  (Correspondingly the median effective temperatures are 3850~K and ~3750~K, respectively.)   This bifurcation is partially reflected in the evolutionary tracks; we see for the 25$M_\odot$ track a short loop that results in increased time spent in region occupied by the early K-type supergiants.  To quantify this further,  in Table~\ref{tab:rsglifetimes} we compute the amount
of time the evolutionary tracks predict a star will spend in the RSG regime ($<$4300~K) as a function of effective temperature.  (The ``gap" occurs around 4100-4150~K.) in effective temperature range 4100-4350 and 3600-4100~K.
We see that the gap (which occurs around 4100-4150~K) actually occurs in the 25$M_\odot$ track.  Although the gap is not present in the 15 or 20$M_\odot$ we will note that the percentage of ``warm" vs.\ ``cool" RSGs does increase with increasing luminosity, in much the same way as is reflected by the quantity of stars seen in the HRD.  A more exact treatment comparison will be made when the higher metallicity tracks become available. 

The numbers in Table~\ref{tab:rsglifetimes} do reveal that the 25$M_\odot$ track may not  extend quite far enough to cooler temperatures, although this may well be an artifact of the metallicity of the solar track being too low for the M31 RSGs; we expect the Hayashi limit\footnote{The Hayashi limit is the coolest temperature at which a star would remain hydrostatically stable \citep{HayashiHoshi}} to shift to cooler temperatures with increasing metallicities \citep[see discussion in][]{EmilyMC}.

\section{Summary and Discussion}
\label{Sec-discussion}

We have shown that the photometric separation of RSGs and red foreground stars using a {\it B-V} vs {\it V-R} diagram works extremely well, at least at the $1.5\times$ solar metallicity of M31.  The exceptions were stars whose photometry was suspect, or marginally on the borderline between that expected for the two sequences.  There was only one exception, out of 354 stars.  

The agreement between the evolutionary tracks and the locations of RSGs in the HRD is excellent; there is one star, J004520.67+414717.3, whose photometry places it clearly in the RSG catagory and yet its high luminosity ($\log L/L_\odot \sim 5.8$) appears to be inconsistent with what the evolutionary tracks predict; it will be interesting to see if this star is still discrepant when higher metallicity tracks become available.  The bifurcation in the effective temperatures that increases
with increasing luminosities is only partially reflected in the evolutionary tracks.   This is consistent with the small extra loop in the 25$M_\odot$ track but is not indicated
by the other tracks.  At higher luminosities the median spectral type becomes slightly later than at lower luminosities; i.e., higher mass RSGs spend more time as cooler temperatures than do stars of lower mass. 

\subsection{Completeness of the Current Sample}
Using the same LGGS photometry that we employ here, \citet{MasseySilva} identified 437 stars as likely RSGs based on the two-color diagrams.   Of these, we have new spectra and spectral types for 255 (about 60\%) of these stars.
The \citet{MasseySilva} study determined radial velocities for 124 of the 437 stars in the original sample.  We have 
new observations of 74 of those 124 stars, and have included the 50 additional stars with radial velocities from \citet{MasseySilva} in Figure~\ref{fig:rvplot}.   We successfully compute physical properties for 192 stars from these
255 stars; the others are either lacking 2MASS K-band photometry or we were unable to fit their effective temperatures. \citet{MasseySilva} only measured physical properties for 16 stars (their radial velocity spectra were at high dispersion and did not provide the necessary amount of wavelength coverage); we have new observations for 8 of those stars here, and have compared the derived properties in Table~\ref{tab:comparison}.  We have chosen to ignore the other 8 stars, as \citet{MasseySilva} inadvertently used solar (rather than $2\times$ solar) models in deriving their physical properties.  Thus Figure~\ref{fig:hrd} and Table~\ref{tab:physical} reflect the physical properties for about 44\% of the original sample.

\subsection{Extension to Fainter Limits}

As discussed in the Introduction, an exciting possibility is to use the luminosity function of RSGs in M31 to provide a constraint on the mass-loss rates of these stars.  In his preliminary work, C. Georgy (2015, private comm.) has shown that the slope of the luminosity function is fairly sensitive to the assumed mass-loss rates.  However, the slopes of the luminosity functions are fairly parallel for $\log L/L_\odot > 4.5$; for this to be useful requires extending our knowledge of the RSG populations down to $\log L/L_\odot\sim 4.2$ or lower.  

\begin{figure}
\epsscale{0.8}
\plotone{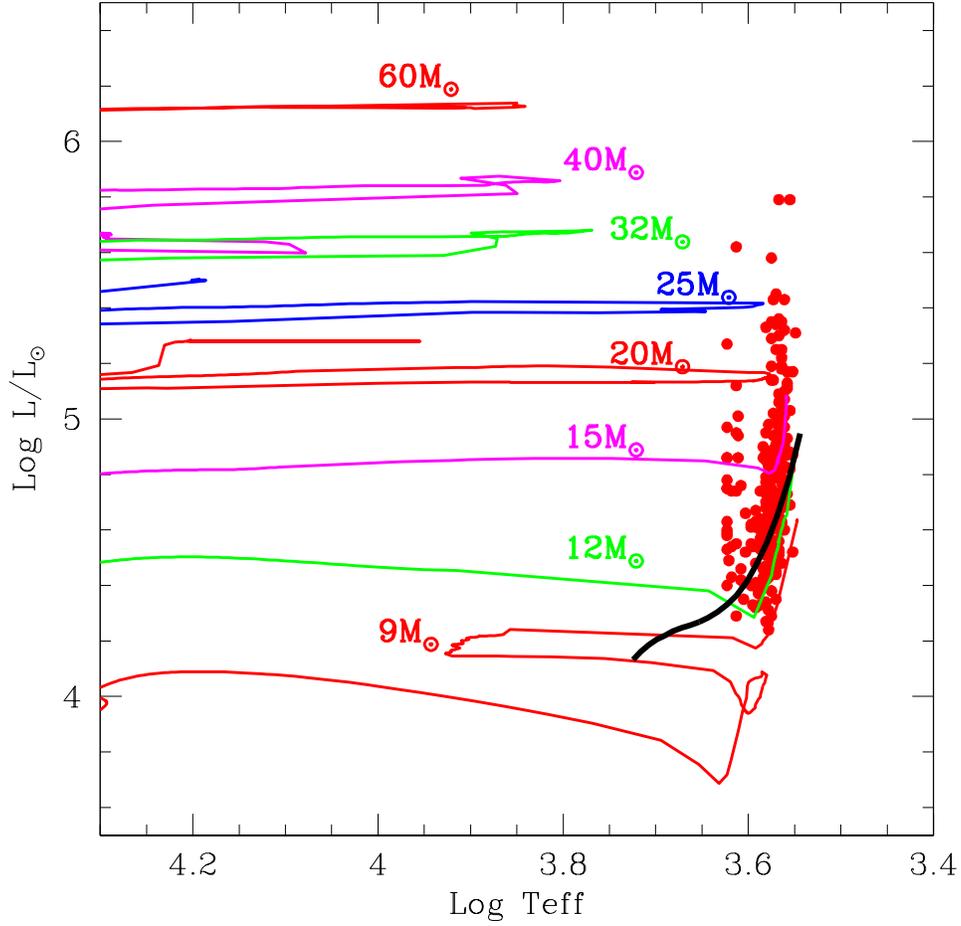}
\epsscale{1.0}
\caption{\label{fig:hrd} Location of the RSGs in the HRD.  We have plotted the location of M31's RSGs in the HRD using the values in Table~\ref{tab:physical}. The evolutionary tracks are from \citet{Sylvia}, computed for solar metallicity and an initial rotation of 40\% of the critical breakup speed.  The initial masses are shown for each track. The black curve corresponds to $V=20,$ our estimated completeness limit.} 
\end{figure}

We include in Figure~\ref{fig:hrd} the current completeness limit set by $V\leq 20.0$ as a solid black line\footnote{There are stars below this line owing to the fact that we used K-band photometry to construct the HRD, and that the $V-band$
brightness of RSGs is variable at the 0.9~mag level or beyond \citep{EmilyVariables}.}.  Including the coolest stars, we would consider the current sample complete to $\log L/L_\odot$ of 4.8.  Thus, we would need to extend this work about 0.6~dex fainter, or about 1.5~mag, to $V\sim 21.5$.  The photometry in the LGGS is good enough to distinguish foreground and RSGs at that level: the expected 1$\sigma$ error in {\it B-V} would be about 0.05 \citep[c.f., Table 6 in][]{LGGSI}, still acceptable (see Figure~\ref{fig:m312col}).  What is lacking, however, is the necessary NIR photometry.  The 2MASS point-source catalog peters out around $K_s\sim 15.0$ in M31; optimally we would need to extend this to $K_s\sim$ 16.5.  It may not be practical to obtain optical spectroscopy at this level; \citet{MasseySilva} notes that 
extending the study just to $V=21.0$ would increase the number of RSG candidates to 1750; we calculate here that
extension to $V=21.5$ would then include $\sim 3200$  stars.

Photometry alone would allow us to fix the effective temperatures well enough to determine the bolometric luminosities.  According to the MARCS models, $\Delta T_{\rm eff}/\Delta (J-K)=-1777$ K mag$^{-1}$ and $\Delta {\rm BC_K}/\Delta (J-K)=1.33$, both extremely linear (RMS of 10K  and 0.005~mag, respectively)  over the RSG effective temperature range.  Thus a S/N of 30 at J and $K_S$ would allow us to determine a $\log L/L_\odot$ of 0.02 dex (about the size of the points in Figure~\ref{fig:hrd}), and provide a temperature precision of 70~K, just a bit worse than 50~K precision we claim for
the spectral fitting.  We have proposed to do exactly this this in the next observing season.
 
\subsection{Concluding Remarks}

\citet{GeorgesRSGs} has emphasized the importance of mass loss during the RSG phase in regards to the further evolution of these stars, and  has raised the alarming possibility that the majority of blue and yellow supergiants we observe are post-RSG objects. Modern estimates of the mass-loss rates $\dot{M}$ are of the order 10$^{-4} M_\odot$~yr$^{-1}$ \citep{vanLoonMdot}, but the total amount of mass lost during the RSG phase as a function of luminosity is poorly constrained observationally for several reasons.  First,  it requires measuring mid-IR excesses for RSGs with known distances; this sample contains only about $\sim$20 RSGs
\citep[see, e.g., Table 3 in][]{vanLoonMdot}.  Second, even so these measurements only tell us the dust production rate; we then have to apply an uncertain dust-to-gas ratio (300-500?). Thirdly, and most importantly,  knowing the mass-loss rates for a scant number of stars does not tell us much about what the time-averaged values actually are, as most of the mass lost during the RSG phase happens during relatively brief outbursts. The star VY CMa may be the most spectacular example, as its circumstellar material suggests that its mass-loss rate was a factor of $\sim$20 higher in the past than it is at present \citep{SmithVYCMa,DecinVYCMa}.

Extension of our work here to fainter limits now has the potential of allowing us to measure the time-average mass-loss rates using the luminosity functions of these objects.  We have established the our photometric technique is sufficient to distinguish RSGs from foreground stars, and combining this technique with NIR photometry has the potential for addressing this in a novel new way, as outlined by \citet{GeorgesRSGs} and \citet{CyrilIAU}.  A more complete knowledge of the RSG population of M31 should allow us to answer this intriguing question.

\acknowledgements

\vskip -10pt
  
We are grateful to the Steward Observatory Time Allocation Committee  for their generous allocation of observing time on the MMT,  and to Perry Berlind, Mike Calkins, and Marc Lacasse for their excellent support of Hectospec.  P. M. would like to especially thank Calkins for organizing and graciously hosting a very pleasant Thanksgiving dinner for all of the Mt Hopkins observers during the November observing run. Nelson Caldwell managed the difficult task of queue scheduling our program.  Collaborator Kathryn Neugent participated
in these observations and encouraged devoting the ``spare" Hectospec fibers for this additional project; she was also kind enough to make useful comments on a draft on the manuscript, as did Cyril Georgy,  Georges Meynet, and Sylvia Esktr\"{o}m.  An anonymous referee provided constructive comments.  Emily Levesque and Joe Llama provided useful help with some of the coding for fitting the MARCS models. This publication makes use of data products from the Two Micron All Sky Survey, which is a joint project of the University of Massachusetts and the Infrared Processing and Analysis Center/California Institute of Technology, funded by the National Aeronautics and Space Administration and the National Science Foundation (NSF).  K. A. E.'s work was supported through the NSF's Research Experiences for Undergraduates program through Northern Arizona University and Lowell Observatory (AST-1461200), and P. M.'s were partially supported by the NSF through AST-1008020 and by Lowell Observatory.

{\it Facilities:} \facility{MMT(Hectospec)}

\clearpage

\bibliographystyle{apj}
\bibliography{masterbib}

\begin{thebibliography}{}
\expandafter\ifx\csname natexlab\endcsname\relax\def\natexlab#1{#1}\fi

\bibitem[{{Bessell} {et~al.}(1998){Bessell}, {Castelli}, \& {Plez}}]{Bessell}
{Bessell}, M.~S., {Castelli}, F., \& {Plez}, B. 1998, \aap, 333, 231

\bibitem[{{Carpenter}(2001)}]{Carpenter}
{Carpenter}, J.~M. 2001, \aj, 121, 2851

\bibitem[{{Cutri} {et~al.}(2003){Cutri}, {Skrutskie}, {van Dyk}, {Beichman},
  {Carpenter}, {Chester}, {Cambresy}, {Evans}, {Fowler}, {Gizis}, {Howard},
  {Huchra}, {Jarrett}, {Kopan}, {Kirkpatrick}, {Light}, {Marsh}, {McCallon},
  {Schneider}, {Stiening}, {Sykes}, {Weinberg}, {Wheaton}, {Wheelock}, \&
  {Zacarias}}]{2MASS}
{Cutri}, R.~M., {Skrutskie}, M.~F., {van Dyk}, S., {et~al.} 2003, VizieR Online
  Data Catalog, 2246, 0

\bibitem[{{Decin} {et~al.}(2006){Decin}, {Hony}, {de Koter}, {Justtanont},
  {Tielens}, \& {Waters}}]{DecinVYCMa}
{Decin}, L., {Hony}, S., {de Koter}, A., {et~al.} 2006, \aap, 456, 549

\bibitem[{{Drout} {et~al.}(2009){Drout}, {Massey}, {Meynet}, {Tokarz}, \&
  {Caldwell}}]{DroutM31}
{Drout}, M.~R., {Massey}, P., {Meynet}, G., {Tokarz}, S., \& {Caldwell}, N.
  2009, \apj, 703, 441

\bibitem[{{Ekstr{\"o}m} {et~al.}(2012){Ekstr{\"o}m}, {Georgy}, {Eggenberger},
  {Meynet}, {Mowlavi}, {Wyttenbach}, {Granada}, {Decressin}, {Hirschi},
  {Frischknecht}, {Charbonnel}, \& {Maeder}}]{Sylvia}
{Ekstr{\"o}m}, S., {Georgy}, C., {Eggenberger}, P., {et~al.} 2012, \aap, 537,
  A146

\bibitem[{{Evans} \& {Massey}(2015)}]{KateRunaway}
{Evans}, K.~A., \& {Massey}, P. 2015, \aj, 150, 149

\bibitem[{{Fabricant} {et~al.}(2005){Fabricant}, {Fata}, {Roll}, {Hertz},
  {Caldwell}, {Gauron}, {Geary}, {McLeod}, {Szentgyorgyi}, {Zajac}, {Kurtz},
  {Barberis}, {Bergner}, {Brown}, {Conroy}, {Eng}, {Geller}, {Goddard},
  {Honsa}, {Mueller}, {Mink}, {Ordway}, {Tokarz}, {Woods}, {Wyatt}, {Epps}, \&
  {Dell'Antonio}}]{Hecto}
{Fabricant}, D., {Fata}, R., {Roll}, J., {et~al.} 2005, \pasp, 117, 1411

\bibitem[{{Fabricant} {et~al.}(2008){Fabricant}, {Kurtz}, {Geller}, {Caldwell},
  {Woods}, \& {Dell'Antonio}}]{FluxHecto}
{Fabricant}, D.~G., {Kurtz}, M.~J., {Geller}, M.~J., {et~al.} 2008, \pasp, 120,
  1222

\bibitem[{{Fujii} \& {Portegies Zwart}(2011)}]{fujii}
{Fujii}, M.~S., \& {Portegies Zwart}, S. 2011, Science, 334, 1380

\bibitem[{{Georgy} {et~al.}(2015){Georgy}, {Ekstr{\"o}m}, {Hirschi}, {Meynet},
  {Massey}, \& {Groh}}]{CyrilIAU}
{Georgy}, C., {Ekstr{\"o}m}, S., {Hirschi}, R., {et~al.} 2015, IAU General
  Assembly, 22, 2256831

\bibitem[{{Georgy} {et~al.}(2013){Georgy}, {Ekstr{\"o}m}, {Eggenberger},
  {Meynet}, {Haemmerl{\'e}}, {Maeder}, {Granada}, {Groh}, {Hirschi}, {Mowlavi},
  {Yusof}, {Charbonnel}, {Decressin}, \& {Barblan}}]{Cyril002}
{Georgy}, C., {Ekstr{\"o}m}, S., {Eggenberger}, P., {et~al.} 2013, \aap, 558,
  A103

\bibitem[{{Gies} \& {Bolton}(1986)}]{Gies}
{Gies}, D.~R., \& {Bolton}, C.~T. 1986, \apjs, 61, 419

\bibitem[{{Hayashi} \& {Hoshi}(1961)}]{HayashiHoshi}
{Hayashi}, C., \& {Hoshi}, R. 1961, \pasj, 13, 442

\bibitem[{{Hurley-Keller} {et~al.}(2004){Hurley-Keller}, {Morrison}, {Harding},
  \& {Jacoby}}]{Hurley}
{Hurley-Keller}, D., {Morrison}, H.~L., {Harding}, P., \& {Jacoby}, G.~H. 2004,
  \apj, 616, 804

\bibitem[{{Kippenhahn} \& {Weigert}(1990)}]{Kipp}
{Kippenhahn}, R., \& {Weigert}, A. 1990, {Stellar Structure and Evolution}
  (Berlin: Springer-Verlag), 192

\bibitem[{{Levesque} \& {Massey}(2012)}]{EmilyWLM}
{Levesque}, E.~M., \& {Massey}, P. 2012, \aj, 144, 2

\bibitem[{{Levesque} {et~al.}(2007){Levesque}, {Massey}, {Olsen}, \&
  {Plez}}]{EmilyVariables}
{Levesque}, E.~M., {Massey}, P., {Olsen}, K.~A.~G., \& {Plez}, B. 2007, \apj,
  667, 202

\bibitem[{{Levesque} {et~al.}(2005){Levesque}, {Massey}, {Olsen}, {Plez},
  {Josselin}, {Maeder}, \& {Meynet}}]{EmilyMW}
{Levesque}, E.~M., {Massey}, P., {Olsen}, K.~A.~G., {et~al.} 2005, \apj, 628,
  973

\bibitem[{{Levesque} {et~al.}(2006){Levesque}, {Massey}, {Olsen}, {Plez},
  {Meynet}, \& {Maeder}}]{EmilyMC}
---. 2006, \apj, 645, 1102

\bibitem[{{Maeder} {et~al.}(1980){Maeder}, {Lequeux}, \&
  {Azzopardi}}]{Maeder80}
{Maeder}, A., {Lequeux}, J., \& {Azzopardi}, M. 1980, \aap, 90, L17

\bibitem[{{Massey}(1998)}]{MasseyRSGs}
{Massey}, P. 1998, \apj, 501, 153

\bibitem[{{Massey}(2006)}]{M31PCyg}
---. 2006, \apjl, 638, L93

\bibitem[{{Massey}(2013)}]{MasseyRev13}
---. 2013, \nar, 57, 14

\bibitem[{{Massey} {et~al.}(2007{\natexlab{a}}){Massey}, {McNeill}, {Olsen},
  {Hodge}, {Blaha}, {Jacoby}, {Smith}, \& {Strong}}]{LGGSIII}
{Massey}, P., {McNeill}, R.~T., {Olsen}, K.~A.~G., {et~al.} 2007{\natexlab{a}},
  \aj, 134, 2474

\bibitem[{{Massey} {et~al.}(2016){Massey}, {Neugent}, \& {Smart}}]{BigTable}
{Massey}, P., {Neugent}, K., \& {Smart}, B. 2016, \aj, in press

\bibitem[{{Massey} {et~al.}(2007{\natexlab{b}}){Massey}, {Olsen}, {Hodge},
  {Jacoby}, {McNeill}, {Smith}, \& {Strong}}]{LGGSII}
{Massey}, P., {Olsen}, K.~A.~G., {Hodge}, P.~W., {et~al.} 2007{\natexlab{b}},
  \aj, 133, 2393

\bibitem[{{Massey} {et~al.}(2006){Massey}, {Olsen}, {Hodge}, {Strong},
  {Jacoby}, {Schlingman}, \& {Smith}}]{LGGSI}
---. 2006, \aj, 131, 2478

\bibitem[{{Massey} {et~al.}(2005){Massey}, {Plez}, {Levesque}, {Olsen},
  {Clayton}, \& {Josselin}}]{Smoke}
{Massey}, P., {Plez}, B., {Levesque}, E.~M., {et~al.} 2005, \apj, 634, 1286

\bibitem[{{Massey} {et~al.}(2009){Massey}, {Silva}, {Levesque}, {Plez},
  {Olsen}, {Clayton}, {Meynet}, \& {Maeder}}]{MasseySilva}
{Massey}, P., {Silva}, D.~R., {Levesque}, E.~M., {et~al.} 2009, \apj, 703, 420

\bibitem[{{Meynet} {et~al.}(2015){Meynet}, {Chomienne}, {Ekstr{\"o}m},
  {Georgy}, {Granada}, {Groh}, {Maeder}, {Eggenberger}, {Levesque}, \&
  {Massey}}]{GeorgesRSGs}
{Meynet}, G., {Chomienne}, V., {Ekstr{\"o}m}, S., {et~al.} 2015, \aap, 575, A60

\bibitem[{{Neugent} \& {Massey}(2014)}]{NeugentBinaries}
{Neugent}, K.~F., \& {Massey}, P. 2014, \apj, 789, 10

\bibitem[{{Neugent} {et~al.}(2012{\natexlab{a}}){Neugent}, {Massey}, \&
  {Georgy}}]{NeugentM31}
{Neugent}, K.~F., {Massey}, P., \& {Georgy}, C. 2012{\natexlab{a}}, \apj, 759,
  11

\bibitem[{{Neugent} {et~al.}(2010){Neugent}, {Massey}, {Skiff}, {Drout},
  {Meynet}, \& {Olsen}}]{NeugentSMC}
{Neugent}, K.~F., {Massey}, P., {Skiff}, B., {et~al.} 2010, \apj, 719, 1784

\bibitem[{{Neugent} {et~al.}(2012{\natexlab{b}}){Neugent}, {Massey}, {Skiff},
  \& {Meynet}}]{NeugentLMC}
{Neugent}, K.~F., {Massey}, P., {Skiff}, B., \& {Meynet}, G.
  2012{\natexlab{b}}, \apj, 749, 177

\bibitem[{{Sanders} {et~al.}(2012){Sanders}, {Caldwell}, {McDowell}, \&
  {Harding}}]{Sanders}
{Sanders}, N.~E., {Caldwell}, N., {McDowell}, J., \& {Harding}, P. 2012, \apj,
  758, 133

\bibitem[{{Schlegel} {et~al.}(1998){Schlegel}, {Finkbeiner}, \& {Davis}}]{Sch}
{Schlegel}, D.~J., {Finkbeiner}, D.~P., \& {Davis}, M. 1998, \apj, 500, 525

\bibitem[{{Smartt} {et~al.}(2009){Smartt}, {Eldridge}, {Crockett}, \&
  {Maund}}]{SmarttIIP}
{Smartt}, S.~J., {Eldridge}, J.~J., {Crockett}, R.~M., \& {Maund}, J.~R. 2009,
  \mnras, 395, 1409

\bibitem[{{Smith} {et~al.}(2001){Smith}, {Humphreys}, {Davidson}, {Gehrz},
  {Schuster}, \& {Krautter}}]{SmithVYCMa}
{Smith}, N., {Humphreys}, R.~M., {Davidson}, K., {et~al.} 2001, \aj, 121, 1111

\bibitem[{{Sofue} \& {Kato}(1981)}]{M312D}
{Sofue}, Y., \& {Kato}, T. 1981, \pasj, 33, 449

\bibitem[{{Tonry} \& {Davis}(1979)}]{TonryDavis}
{Tonry}, J., \& {Davis}, M. 1979, \aj, 84, 1511

\bibitem[{{van den Bergh}(2000)}]{vandenbergh2000}
{van den Bergh}, S. 2000, {The Galaxies of the Local Group} (Cambridge:
  Cambridge Univ.\ Press)

\bibitem[{{van Loon} {et~al.}(2005){van Loon}, {Cioni}, {Zijlstra}, \&
  {Loup}}]{vanLoonMdot}
{van Loon}, J.~T., {Cioni}, M.-R.~L., {Zijlstra}, A.~A., \& {Loup}, C. 2005,
  \aap, 438, 273

\bibitem[{{Zaritsky} {et~al.}(1994){Zaritsky}, {Kennicutt}, \&
  {Huchra}}]{Zaritsky}
{Zaritsky}, D., {Kennicutt}, Jr., R.~C., \& {Huchra}, J.~P. 1994, \apj, 420, 87

\end{thebibliography}

\clearpage

\begin{deluxetable}{l r r r r r r r l}
\tablecaption{\label{tab:RVs} Radial Velocities RSG and Foreground Candidates}
\tablewidth{0pt}
\tablehead{
\colhead{Star}
&\colhead{$X/R$\tablenotemark{a} }
&\colhead{$V_{\rm M31}$\tablenotemark{b} }
&\colhead{$V_{\rm obs}$\tablenotemark{b} }
&\colhead{$V_{\rm err}$\tablenotemark{b} }
&\colhead{$N_{\rm obs}$}
&\colhead{TDR\tablenotemark{c} }
&\colhead{Old $V_{\rm obs}$\tablenotemark{b,}\tablenotemark{d}}
&\colhead{Phot.\ Type\tablenotemark{e} }
}
\startdata
J003950.86+405332.0&-0.586&-436.5&-433.0&    0.5& 6& 14.4&\nodata&RSG\\
J003950.98+405422.5&-0.563&-431.0&-418.4&    0.7& 3& 14.5&\nodata&RSG\\
J003957.00+410114.6&-0.409&-393.8&-446.9&    0.6& 6& 12.7& -442.4&RSG\\
J004015.18+405947.7&-0.491&-413.5&-411.5&    0.5& 6& 12.9& -406.8&RSG\\
J004015.86+405514.1&-0.638&-449.0&-387.9&    0.9& 4&  8.7&\nodata&RSG\\
J004019.15+404150.8&-0.999&-536.2&-564.6&    1.2& 1& 13.4& -551.9&RSG\\
J004020.06+410651.3&-0.326&-373.8&-362.6&    0.5& 5& 15.8&\nodata&RSG\\
J004023.84+410458.6&-0.375&-385.5&-374.1&    1.2& 2&  9.0&\nodata&RSG\\
J004025.36+404623.1&-0.967&-528.6&-565.2&    0.6& 6& 10.6&\nodata&RSG\\
\enddata
\tablecomments{Table~\ref{tab:RVs} is published in its entirety in a machine readable format. A portion is shown here for guidance regarding its form and content.}
\tablenotetext{a}{$X$ is the distance along the semi-major axis, and $R$ is the galactocentric radius within the
plane of M31.}
\tablenotetext{b}{Radial velocities in km s$^{-1}$}
\tablenotetext{c}{\citet{TonryDavis} $r$ parameter.}
\tablenotetext{d}{From \citet{MasseySilva}.}
\tablenotetext{e}{Photometric type based upon {\it B-V} vs {\it V-R}; fgd = foreground; RSG = red supergiant.}
\tablenotetext{f}{Radial velocity disagrees with photometric type.}
\end{deluxetable}

\begin{deluxetable}{l l c r c r r r}
\tablecaption{\label{tab:physical} Physical Properties}
\tablewidth{0pt}
\tablehead{
&&&\multicolumn{2}{c}{$K$-band Photom.} &&  \\ \cline{4-5}
\colhead{Star}
&\colhead{Type}
&\colhead{$T_{\rm eff}$}
& \colhead{$K_s$}
&\colhead{Source\tablenotemark{a}}
&\colhead{$M_K$}
&\colhead{$\log L/L_\odot$}
&\colhead{$R/R_\odot$}
}
\startdata
J003950.86+405332.0 & M1 I     &   3850  &   14.45  & M &  -10.03 &    4.86 &     600\\
J003950.98+405422.5 & M4 I     &   3650  &   14.20  & M &  -10.28 &    4.89 &     700\\
J003957.00+410114.6 & M4 I     &   3650  &   15.08  & F &   -9.40 &    4.54 &     470\\
J004015.18+405947.7 & M3 I     &   3700  &   14.26  & M &  -10.22 &    4.89 &     670\\
J004015.86+405514.1 & K2 I     &   3950  &   15.62  & M &   -8.86 &    4.42 &     340\\
J004019.15+404150.8 & M2 I     &   3750  &   14.61  & M &   -9.87 &    4.76 &     570\\
J004020.06+410651.3 & M0 I     &   3900  &   14.60  & M &   -9.88 &    4.81 &     560\\
J004023.84+410458.6 & K5 I     &   3975  &   15.69: & M &   -8.79 &    4.40 &     330\\
J004025.36+404623.1 & M0 I     &   3900  &   15.29: & M &   -9.19 &    4.53 &     400\\
J004025.75+404254.8 & K5 I     &   3925  & \nodata  & X & \nodata & \nodata & \nodata\\
\enddata
\tablecomments{Table~\ref{tab:physical} is published in its entirety in a machine readable format. A portion is shown here for guidance regarding its form and content.}
\tablenotetext{a}{M=2MASS \citep{2MASS}, F=FLAMINGOS \citep{MasseySilva}, X=no data.}
\end{deluxetable}

\begin{deluxetable}{l l l l c c}
\tablecaption{\label{tab:comparison} Comparison with \citet{MasseySilva}}
\tablewidth{0pt}
\tablehead{
&
\multicolumn{2}{c}{Spectral Type} 
&
&\multicolumn{2}{c}{Effective Temperature} \\ \cline{2-3} \cline{5-6}
\colhead{Star}
&\colhead{New}
&\colhead{Old}
&
&\colhead{New}
&\colhead{Old\tablenotemark{a}}
}
\startdata
J003957.00+410114.6&M4 I   &M0 I&&3650& 3775\\
J004035.08+404522.3&M2 I   &M2.5 I&&3750& 3775\\
J004047.82+410936.4&M3 I   &M3 I&&3650& 3725\\
J004124.80+411634.7&M2 I   &M3+? I&&3725& 3700\\
J004255.95+404857.5&M2 I   &M2 I&&3800& 3750\\
J004428.71+420601.6&M0 I   &M0  I&&3825& 3900\\
J004454.38+412441.6&M0 I   &M2  I&&3850& 3800\\
J004514.95+414625.6&M1 I   &M2  I&&3800& 3750\\
\enddata
\tablenotetext{a}{Adjusted by +75 K; see text.}
\end{deluxetable}

\begin{deluxetable}{l r r r r r r r}
\tabletypesize{\scriptsize}
\tablecaption{\label{tab:rsglifetimes} Lifetimes (years) for RSGs from the \citet{Sylvia} Evolutionary Tracks}
\tablewidth{0pt}
\tablehead{
\colhead{Initial mass}
& \multicolumn{6}{c}{Effective temperatures (K)}
& \colhead{``warm" / ``cool"\tablenotemark{a}}  \\ \cline{2-7}
\colhead{$(M_\odot)$}
&\colhead{3700-3800}
&\colhead{3800-3900}
&\colhead{3900-4000}
&\colhead{4000-4100}
&\colhead{4100-4200}
&\colhead{4200-4300}
}
\startdata
25 &      0       & 63,000 & 30,00 & 8,00 & 3,000 & 12,000 & 15\% \\
20 & 151,000 & 149,000 & 20,000 & 6,000 & 4,300 & 5,400 & 3\% \\
15 & 380,000 & 8,000 & 3,000 & 2,000 & 2,000 & 2,000 & 0.3!\% \\
\enddata
\tablenotetext{a}{``Warm" is 4100-4300~K while ``cool" is $<4100$~K.}
\end{deluxetable}
\end{document}